\documentclass[12pt]{article}

\input epsf
\usepackage{amssymb}
\usepackage[dvips]{graphicx}
\usepackage{amsmath,amscd}

\def\dj{\hbox{d\kern-0.347em \vrule width 0.3em height 1.252ex depth
-1.21ex \kern 0.051em}}

\begin{document}

\setlength{\oddsidemargin}{0cm}
\setlength{\baselineskip}{7mm}


\thispagestyle{empty}
\setcounter{page}{0}

\begin{flushright}

\end{flushright}

\vspace*{1cm}

\begin{center}
{\bf \Large A Note on the Collision of Reissner-Nordstr\"om}

\vspace*{0.5cm}

{\bf \Large Gravitational Shock Waves in AdS}

\vspace*{0.3cm}

\vspace*{1cm}

\'Alvaro Due\~nas-Vidal\footnote{\tt 
adv@usal.es}
and Miguel \'A. V\'azquez-Mozo\footnote{\tt 
Miguel.Vazquez-Mozo@cern.ch}

\end{center}

\vspace*{0.0cm}

\begin{center}
  
 {\sl Departamento de F\'{\i}sica Fundamental and IUFFyM, \\
 Universidad de Salamanca \\ 
 Plaza de la Merced s/n,
 E-37008 Salamanca, Spain
  }

\end{center}

\vspace*{2.cm}

\centerline{\bf \large Abstract}

We study the collision of two Reissner-Nordstr\"om gravitational shock waves in AdS and show that 
the charge completely prevents the formation of marginally trapped surfaces of the Penrose type with
topology $S^{D-2}$, independently of the energy and the value of the impact parameter.
In the case of head-on collisions, a numerical analysis shows that no trapped surfaces with topology $S^{1}
\times S^{D-3}$ form either.  

\newpage

\paragraph{Introduction.}

The study of collision of gravitational waves has been a very active area of research in the field of general 
relativity (see \cite{griffiths} for a summary of results). The interest in the subject has been renewed in the
context of the AdS/CFT correspondence \cite{ads/cft}. The underlying reason is the expectation that
colliding waves in AdS spacetime could provide a reliable gravitational dual of the high-energy 
collision of two ``nuclei'', modeled by energy lumps in 
the holographic strongly coupled gauge theory \cite{collisionsAdS,collisionsAdS_general}. 
Although these energy lumps do not reproduce all the properties of heavy ions, 
these theoretical experiments can be used as models where collective gauge theory effects, relevant for realistic heavy ion
collisions, are tractable. An expected outcome of the collisions of two gravitational waves is the formation of a black hole. 
Holographically, black hole formation can be interpreted as the thermalization of the gauge theory plasma resulting from the
collision of the lumps.

In spite of the recent progress in the numerical study of the problem of the collision of two gravitational
waves \cite{AdSGWcolnum}, a full understanding of the highly nonlinear dynamics dominating the physics 
in the interaction region is still missing. An alternative to a full numerical simulation of the collision process (in
the spirit of numerical general relativity) is to look for the formation of a trapped surface that would signal 
the eventual presence of an event horizon \cite{collisionsAdS}. Although in many cases this also involves
numerical analysis, the approach is technically simpler. The idea \cite{flat} consists in looking for marginally
trapped surfaces with topology $S^{D-2}$,
lying on the hypersurface $\{u=0,v<0\}\cup \{u<0,v=0\}$, which does not involve solving for the
geometry in the interaction region. The information about the collision
of the waves is nevertheless encoded in the nontrivial matching condition at $u=v=0$. In the following, we will
refer to these marginally trapped surfaces as of Penrose type.

Varying the parameters of the collision, it is possible to find thresholds for the formation of Penrose
trapped surfaces.
Holographically, this can be seen as a threshold for the onset of the 
plasma thermalization after the collision takes place. This is what happens, for example, for large enough impact parameter \cite{LinShuryak,DVVM1}, 
or when the sizes of the two colliding lumps are very different \cite{DVVM1}. A more intriguing threshold was found in
\cite{AGGSVTVM2} associated with the transverse size of the gravitational source of the shock wave in AdS, which however
does not induce a change in the holographic energy-momentum tensor.

\paragraph{Trapped surfaces with topology $S^{D-3}$ in the collision of RN-AdS shock waves.}

It would be desirable to extend the analysis of the formation of marginally trapped surfaces of Penrose type to
more general types of incoming waves, in particular those obtained taking an infinite boost limit \cite{AS} of a 
$D$-dimensional Reissner-Nordstr\"om (RN) solution asymptotically anti-de Sitter (AdS)
\begin{eqnarray}
ds^{2}&=&-\left[1-{2G_{N}M\over r^{D-3}}+{G_{N}Q^{2}\over r^{2(D-3)}}+{r^{2}\over L^{2}}\right]dt^{2} 
\nonumber \\[0.2cm]
&+&\left[1-{2G_{N}M\over r^{D-3}}+{G_{N}Q^{2}\over r^{2(D-3)}}+{r^{2}\over L^{2}}\right]^{-1}dr^{2}+r^{2}d\Omega_{D-2}^{2},
\end{eqnarray}
where $L$ is the radius of AdS, and $M$ and $Q$ are given in terms of the mass $m$ and electric charge $\mathfrak{q}$ by
\begin{eqnarray}
M={8\pi m\over (D-2)\Omega_{D-2}}, \hspace*{1cm} Q^{2}={8\pi \frak{q}^{2}\over (D-2)(D-3)},
\end{eqnarray}
with $\Omega_{N}$ the volume of the $N$-dimensional sphere $S^{N}$. Performing a boost with Lorentz parameter 
$\gamma$, the shock wave geometry is obtained in the limit $\gamma\rightarrow\infty$ while keeping
\begin{eqnarray}
\mu\equiv \gamma M, \hspace*{1cm} e^{2}\equiv \gamma Q^{2}
\end{eqnarray} 
fixed \cite{lousto_sanchez,yoshino_mann,arefeva_et_al_1}. The metric takes the form (in Poincar\'e coordinates)
\begin{eqnarray}
ds^{2}={L^{2}\over z^{2}}\left[-dudv+dz^{2}+d\vec{x}^{\,2}_{T}+{z\over L}\Phi(q)_{\rm RN}\delta(u)du^{2}\right].
\label{eq:wave_poincare}
\end{eqnarray}
The metric function $\Phi(q)_{\rm RN}$ depends on the AdS chordal coordinate 
\begin{eqnarray}
q={(z-L)^{2}+\vec{x}_{T}^{\,2}\over 4Lz}
\label{eq:chordal_coordinate}
\end{eqnarray}
and can be expressed as
\begin{eqnarray}
\Phi(q)_{\rm RN}=\Phi(q)_{D}-{e^{2}\over 2\mu}\Phi(q)_{2D-3},
\label{eq:phiRN}
\end{eqnarray}
where $\Phi(q)_{D}$ is the profile of the Aichelburg-Sexl shock wave in AdS$_{D}$ \cite{hotta_tanaka,collisionsAdS}. Therefore we have
\begin{eqnarray}
\Phi(q)_{\rm RN}&=&L {2^{3-D}\sqrt{\pi}\Gamma\left({D\over 2}\right)\over
\Gamma\left({D+1\over 2}\right)}\left({G_{N}\mu\over L^{D-3}}\right)
q^{2-D}{}_{2}F_{1}\left(D-2,{D\over 2};D;-{1\over q}\right) \nonumber \\[0.2cm]
&-&
L{2^{5-2D}\sqrt{\pi}\Gamma\left({2D-3\over 2}\right)\over
\Gamma\left(D-1\right)}\left({\sqrt{G_{N}}e\over L^{D-3}}\right)^{2}
q^{5-2D}{}_{2}F_{1}\left(2D-5,{2D-3\over 2};2D-3;-{1\over q}\right). 
\label{eq:PhiRN}
\end{eqnarray}
To avoid confusion, we recall that the energy parameter $\mu=\gamma M$ used here and in \cite{arefeva_et_al_1} 
is related to the 
parameter $E=\gamma m$ of Ref. \cite{collisionsAdS} by the rescaling $E={D-2\over 8\pi}\Omega_{D-2}\mu$.

We analyze the problem of collision of two shock waves. Generically, in the region outside the interaction wedge, the 
metric reads
\begin{eqnarray}
ds^{2}={L^{2}\over z^{2}}\left[-dudv+dz^{2}+d\vec{x}_{\rm T}^{\,2}+{z\over L}\Phi_{-}(z,\vec{x}_{T})\delta(u)du^{2}
+{z\over L}\Phi_{+}(z,\vec{x}_{T})\delta(v)dv^{2}\right].
\end{eqnarray}
It is convenient to change coordinates $(u,v,z,\vec{x}_{T})\Rightarrow (U,V,Z,\vec{X}_{T})$
to remove the distributional terms in the metric of the incoming waves
(see the Appendix of Ref. \cite{DVVM1}). Following
Penrose \cite{flat}, we look for 
surfaces $\mathcal{S}$ 
with topology $S^{D-2}$ contained in the past ligh-cone $\{U\leq 0,V=0\}\cup\{U=0,V\leq 0\}$. This is parametrized
in terms of two {\em nonnegative} functions $\psi_{\pm}(Z,\vec{X}_{T})$ as
\begin{eqnarray}
\mathcal{S}_{+}=\left\{
\begin{array}{l}
U=-\psi_{+}(Z,\vec{X}_{T}) \\[0.2cm]
V=0 
\end{array}
\right. 
,\hspace*{1cm}
\mathcal{S}_{-}=\left\{
\begin{array}{l}
U=0 \\[0.2cm]
V=-\psi_{-}(Z,\vec{X}_{T})
\end{array}
\right.
,
\end{eqnarray}
where $\mathcal{S}=\mathcal{S}_{+}\cup \mathcal{S}_{-}$. Defining 
\begin{eqnarray}
\Psi_{\pm}(Z,\vec{X}_{T})={Z\over L}\psi_{\pm}(Z,\vec{X}_{T}),
\end{eqnarray}
the condition that the expansion of the congruence of null geodesics normal to $\mathcal{S}$ vanishes reads
\begin{eqnarray}
\left(\Box_{\mathbb{H}_{D-2}}-{D-2\over L^{2}}\right)(\Phi_{\pm}-\Psi_{\pm})=0,
\label{eq:condition1}
\end{eqnarray}
where $\Phi_{\pm}$ are the profiles of the two incoming waves.

In the case of head-on collisions, we can exploit the O($D-2$) invariance of the system and assume 
that $\Psi_{\pm}(Z,\vec{X}_{T})=\Psi_{\pm}(q)$ only depends on the chordal coordinate \eqref{eq:chordal_coordinate}. 
In terms of it, the metric of the hyperbolic transverse space takes the form
\begin{eqnarray}
ds^{2}_{\mathbb{H}_{D-2}}=L^{2}\left[{dq^{2}\over q(q+1)}+4q(q+1)d\Omega^{2}_{D-3}\right],
\end{eqnarray}
and the trapped surface equation \eqref{eq:condition1} is given by
\begin{eqnarray}
\left[q(q+1){d^{2}\over dq^{2}}+{D-2\over 2}(1+2q){d\over dq}-(D-2)\right]\Big[\Phi_{\pm}(q)-\Psi_{\pm}(q)\Big]=0.
\label{eq:condition1_chordal}
\end{eqnarray}

Since $\mathcal{S}$ has two branches, we have to make sure that the null geodesics are continuous across the $(D-3)$-dimensional
surface $\mathcal{C}=\mathcal{S}_{+}\cap\mathcal{S}_{-}$ defined by $\Psi_{\pm}(q)=0$. This is implemented by the boundary condition
\begin{eqnarray}
g^{ab}\partial_{a}\Psi_{\pm}\partial_{b}\Psi_{\pm}\Bigg|_{\mathcal{C}}=4,
\label{eq:gabpartial=4}
\end{eqnarray} 
where $g_{ab}$ is the metric on $\mathbb{H}_{D-2}$. 
Using the chordal coordinate, the surface $\mathcal{C}$ is parametrized by $q=q_{\mathcal{C}}$, and 
the previous condition can be written as
\begin{eqnarray}
\Psi'_{\pm}(q_{\mathcal{C}})^{2}={4L^{2}\over {q_{\mathcal{C}}(q_{\mathcal{C}}+1)}} \hspace*{0.5cm}
\Longrightarrow\hspace*{0.5cm} \Psi_{\pm}'(q_{\mathcal{C}})=-{2L\over \sqrt{q_{\mathcal{C}}(q_{\mathcal{C}}+1)}}.
\end{eqnarray}
The sign is fixed by the requirement that the trapped surface has topology $S^{D-2}$, which means that $\Psi_{\pm}(q)>0$ for $0\leq q
<q_{\mathcal{C}}$. The trapped surface is determined by the solution to the differential equation \eqref{eq:condition1_chordal}
subjected to the two boundary conditions 
\begin{eqnarray}
\Psi_{\pm}(q_{\mathcal{C}}) &=& 0 \nonumber \\
\Psi_{\pm}'(q_{\mathcal{C}})&=&-{2L\over \sqrt{q_{\mathcal{C}}(q_{\mathcal{C}}+1)}}
\label{eq:boundary_cond}
\end{eqnarray}

For simplicity, let us assume collision between two identical shock waves, so we have $\Psi_{+}(q)=\Psi_{-}(q)\equiv \Psi(q)$
and $\Phi_{+}(q)=\Phi_{-}(q)\equiv \Phi(q)_{\rm RN}$. 
The general solution to the differential equation \eqref{eq:condition1_chordal} can be written
\begin{eqnarray}
\Psi(q)=\Phi(q)_{\rm RN}+C_{1}\Phi_{1}(q)+C_{2}\Phi_{2}(q),
\label{eq:generalsolutiondiffeq}
\end{eqnarray}
where $C_{1,2}$ are two integration constants and 
$\Phi_{1,2}(q)$ are the two independent solutions to the homogeneous differential equation
\begin{eqnarray}
\Phi_{1}(q) &=& 1+2q \nonumber \\
\Phi_{2}(q) &=& q^{2-D}{}_{2}F_{1}\left(D-2,{D\over 2};D;-{1\over q}\right).
\end{eqnarray}
If the trapped surface has topology $S^{D-2}$, the function $\Psi(q)$ has to be regular for $0\leq q\leq q_{\mathcal{C}}$ and this
forces $C_{2}=0$. Imposing besides $\Phi_{1}(q_{\mathcal{C}})=0$ determines the remaining constant to be
\begin{eqnarray}
C_{1}=-{1\over 1+2q_{\mathcal{C}}}\Phi(q_{\mathcal{C}})_{\rm RN}. 
\end{eqnarray}
The value of $q_{\mathcal{C}}$ is now determined by the second condition \eqref{eq:boundary_cond}. Plugging
\begin{eqnarray}
\Psi(q)=\Phi(q)_{\rm RN}-{1+2q\over 1+2q_{\mathcal{C}}}\Phi(q_{\mathcal{C}})_{\rm RN},
\label{eq:Psi(q)withcharge}
\end{eqnarray}
into the second equation in \eqref{eq:boundary_cond}, we find the algebraic equation
\begin{eqnarray}
\Phi'(q_{\mathcal{C}})_{\rm RN}-{2\over 1+2q_{\mathcal{C}}}\Phi(q_{\mathcal{C}})_{\rm RN}+
{2L\over \sqrt{q_{\mathcal{C}}(q_{\mathcal{C}}+1)}}=0.
\label{eq:contcondtrivtop}
\end{eqnarray}

\begin{figure}
\centerline{\includegraphics[width=2.8in]{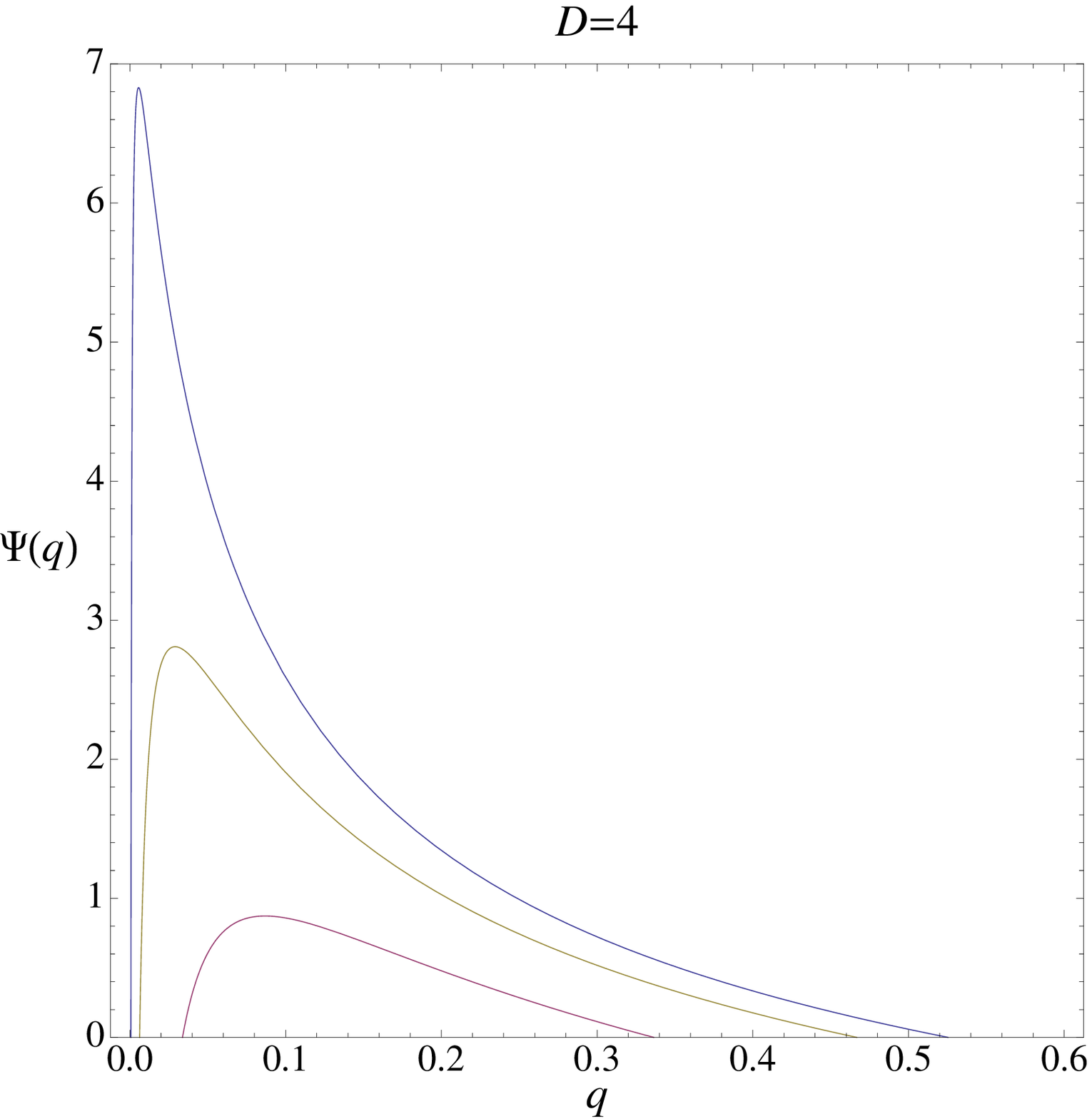}\hspace*{1cm}
\includegraphics[width=2.8in]{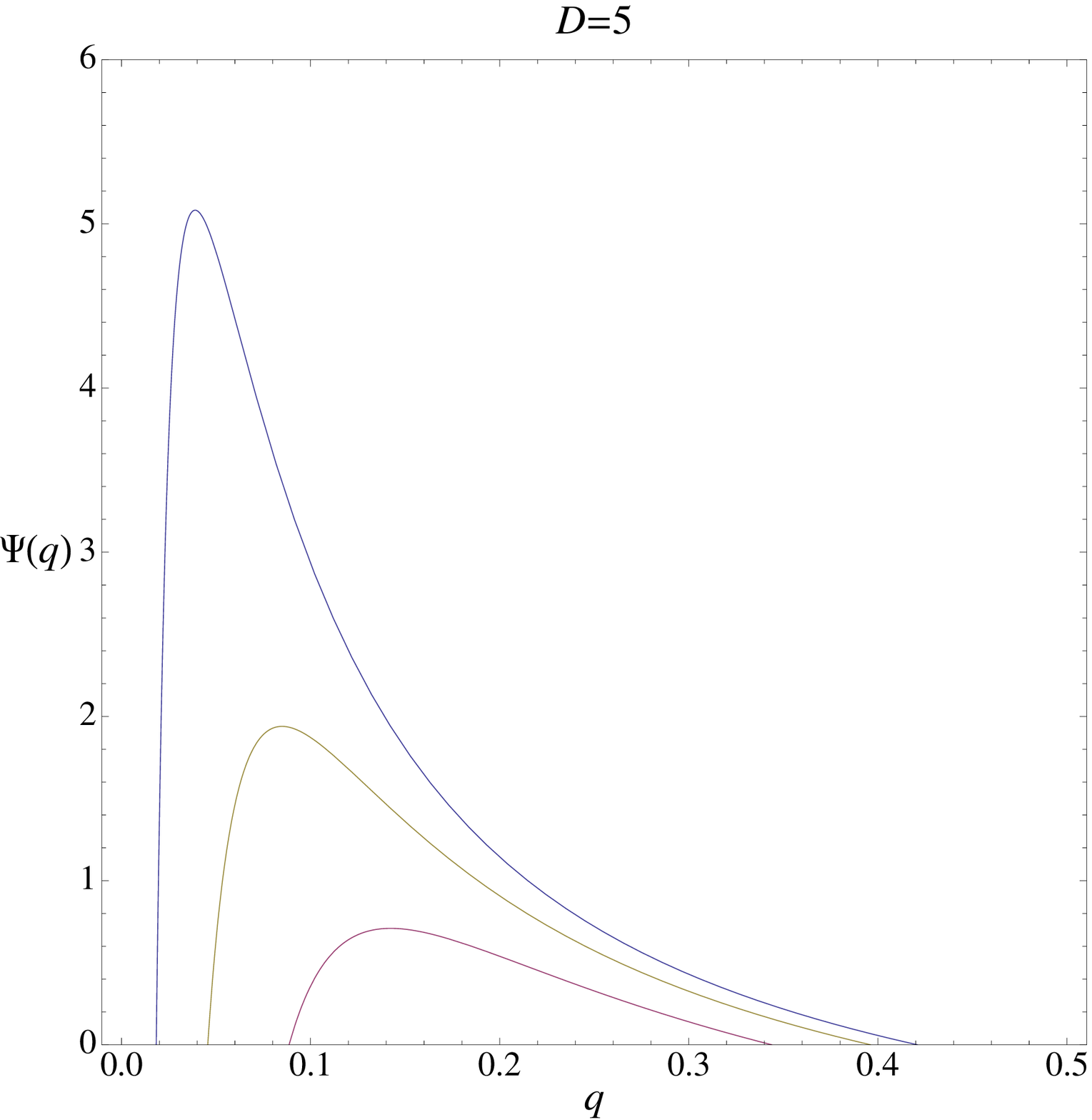}}
\caption{Plot of the function $\Psi(q)$ for $D=4$ (left panel) and $D=5$ (right panel). In all cases the energy of the 
head-on collision is $G_{N}\mu/L^{D-3}=1$ and the charge parameter (from top to bottom) $\sqrt{G_{N}}e/L^{D-3}=0.5$, $0.75$ and 
$1.0$.}
\label{fig1_head_on}
\end{figure}

The problem of the existence of trapped surfaces of the Penrose type in the collision of two RN-AdS shock waves has
been studied in \cite{arefeva_et_al_1} (see also \cite{arefeva_et_al_2}), 
as well as in a first version of this paper. The analysis presented
there, however, has a fundamental 
flaw
consisting in that the function $\Psi(q)$ has a second zero 
below $q_{\mathcal{C}}$ and therefore the surface defined by it does not have topology $S^{D-2}$ as 
assumed\footnote{We thank an anonymous referee for pointing out this 
basic problem that we overlooked in the previous version of this paper.}.
The origin of this problem lies in the fact that the charge-dependent term in $\Phi_{\rm RN}(q)$ shown in 
Eq. \eqref{eq:PhiRN} tends to minus infinity as $q\rightarrow 0^{+}$.  Indeed, whereas
in this limit 
the $\mu$-dependent term diverges as $q^{4-D\over 2}$ for $D>4$ and $-\log q$ for $D=4$, 
the $e$-dependent term diverges as $-q^{7-2D\over 2}$ for $D\geq 4$. 
This second term 
dominates near $q=0$ for any $e>0$ (and $D\geq 4$), so $\Psi(q)$ goes to minus infinity as $q\rightarrow 0^{+}$
[see Eq. \eqref{eq:Psi(q)withcharge}].
This behavior is explicitly shown in Fig. \ref{fig1_head_on}, where the
data for the function $\Psi(q)$ is plotted for $D=4$ and $D=5$ and various values of the charge. 

On physical grounds it is to expect that this problem remains when considering a nonvanishing impact parameter for the collision. 
To check this requires solving numerically 
the Laplace-type equation \eqref{eq:condition1} with the appropriate boundary conditions
\cite{yoshino_nambu,LinShuryak,DVVM1}. We work in radial coordinates $r=2L\sqrt{q(q+1)}$,
where the 
metric of the transverse hyperbolic space takes the form
\begin{eqnarray}
ds^{2}_{\mathbb{H}_{D-2}}={dr^{2}\over 1+{r^{2}\over L^{2}}}+r^{2}d\theta^{2}+r^{2}\sin^{2}\theta d\Omega_{D-4}^{2}.
\end{eqnarray}
The system is O($D-3$)-invariant and the Penrose trapped 
surface is parametrized by $\Psi_{\pm}(r,\theta)$. To solve for these functions, it is convenient to define
$H_{\pm}(r,\theta)=\Phi_{\pm}(r,\theta)_{\rm RN}-\Psi_{\pm}(r,\theta)$, where $\Phi_{\pm}(r,\theta)_{\rm RN}$ are the 
profiles of the incoming RN-AdS shock 
waves whose sources are located respectively at $r_{\pm}={b\over 2}$, 
$\theta_{+}=0$, $\theta_{-}=\pi$, $\vartheta^{i}_{\pm}=0$, with
$\vartheta^{i}_{\pm}$ the angular coordinates on $S^{D-4}$. 
$H_{\pm}(r,\theta)$ satisfies the equation
\begin{eqnarray}
\left[\left(1+{r^{2}\over L^{2}}\right)\partial_{r}^{2}+{(D-3)L^{2}+(D-2)r^{2}\over rL^{2}}\partial_{r}
+{1\over r^{2}}\partial_{\theta}^{2}+{D-4\over r^{2}\tan\theta}\partial_{\theta}-{D-2\over L^{2}}\right]
H_{\pm}(r,\theta)=0.
\label{eq:laplace-like}
\end{eqnarray}
This has to be solved in a domain $0\leq \theta\leq \pi$, $0\leq r(\theta)\leq L G(\theta)$. As a consequence of the symmetry of the problem, the functions $\Phi_{\pm}(r,\theta)_{\rm RN}$ and $\Psi_{\pm}(r,\theta)$ satisfy
\begin{eqnarray}
\Psi_{\pm}(r,\theta)=\Psi_{\mp}(r,\pi-\theta), \hspace*{1cm}
\Phi_{\pm}(r,\theta)_{\rm RN}=\Phi_{\mp}(r,\pi-\theta)_{\rm RN}.
\end{eqnarray}
This implies Neumann boundary conditions for the function $H_{\pm}(r,\theta)$ on the lower boundary, 
$\partial_{\theta}H_{\pm}(r,0)=0=\partial_{\theta}H_{\pm}(r,\pi)$. 

The function $G(\theta)$ determining the shape of the candidate trapped surface 
also has the symmetry $G(\theta)=G(\pi-\theta)$. It has to be chosen such that the following conditions are satisfied
\begin{eqnarray}
H_{\pm}(r,\theta)\Bigg|_{r=LG(\theta)}&=&\Phi_{\pm}\left(LG(\theta),\theta\right)_{\rm RN}, \nonumber \\
\left.
\left[\left(1+{r^{2}\over L^{2}}\right)(\partial_{r}\Psi_{+})(\partial_{r}\Psi_{-})
+{1\over r^{2}}(\partial_{\theta}\Psi_{+})(\partial_{\theta}\Psi_{-})\right]
\right|_{r=LG(\theta)}&=&4.
\end{eqnarray}
We solve this boundary problem for the collision of two identical shocks using the method devised in \cite{yoshino_nambu}:
we solve Eq. \eqref{eq:laplace-like} numerically in a $50(\mbox{angular})\times  100(\mbox{radial})$ 
grid and find $G(\theta)$ through a
trial-and-correction loop. In the following we summarize our results for the collision of two waves of the same energy.
The details of the implementation of the method can be found in \cite{DVVM1}, where
it was applied to the off-center collision of two Aichelburg-Sexl-AdS shock waves in various dimensions.

\begin{figure}
\centerline{\includegraphics[width=2.8in]{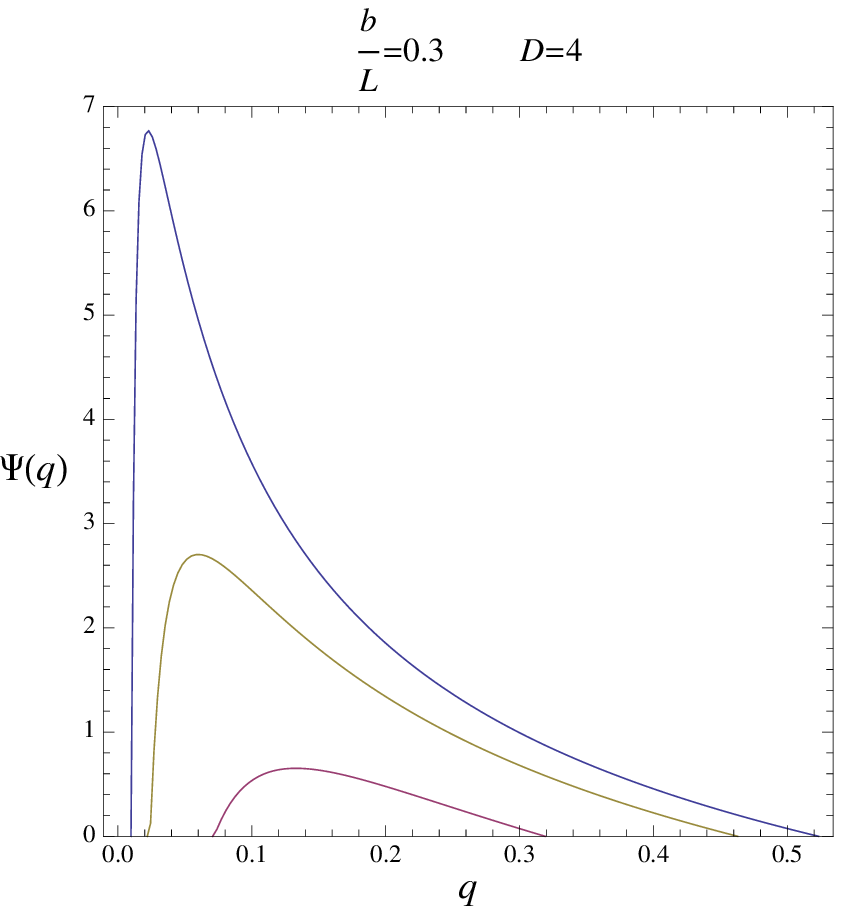}\hspace*{1cm}
\includegraphics[width=2.8in]{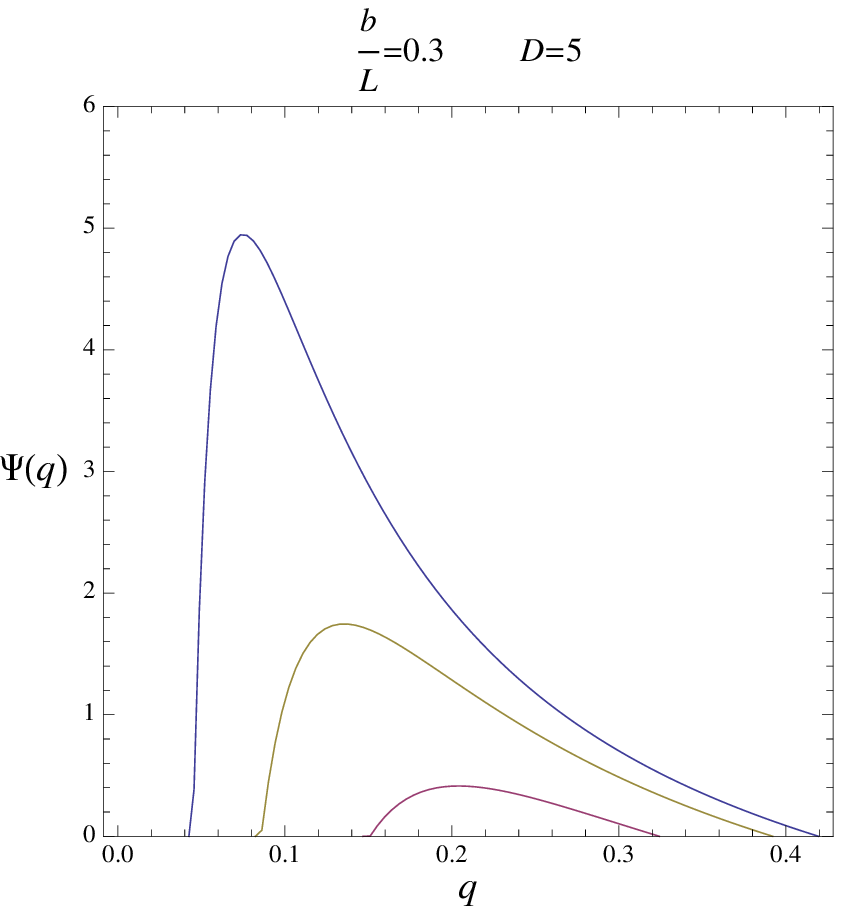}}
\caption{Plot of the section $\theta=0$ of the function $\Psi(q,\theta)$
in four (left panel) and five dimensions (right panel) with nonvanishing impact parameter
$b/L=0.3$. 
Again, the energy of the collision is $G_{N}\mu/L^{D-3}=1$ and the charge parameter (from top to bottom) $\sqrt{G_{N}}e/L^{D-3}=0.5$, $0.75$ and 
$1.0$.}
\label{fig2_impact}
\end{figure}

The numerical data obtained shows that $\Psi(r,\theta)$ not only vanishes at the boundary $r=L G(\theta)$,
but also in the interior. In Fig. \ref{fig2_impact} we have plotted the radial profile of the function $\Psi(r,\theta_{0})$ at
$\theta_{0}=0$, both for $D=4$ (left panel) and $D=5$ (right panel). 
The conclusion is that there are no trapped surfaces of the Penrose type with topology $S^{D-2}$ produced in the 
collision of two RN-AdS shock waves, both with or without impact parameter.

\paragraph{Changing the topology.}

Given the results above, it seems natural to look for marginally trapped 
surfaces with topology $S^{1}\times S^{D-3}$. Mathematically, this means that we allow for the possibility of 
$\Psi(q)$ vanishing at two points that we denote by $q_{\rm in}$ and $q_{\rm out}$, and such that
\begin{eqnarray}
\Psi_{\pm}(q)>0 \hspace*{1cm} \mbox{for} \hspace*{1cm} q_{\rm in}<q<q_{\rm out}. 
\label{eq:psi>0interm}
\end{eqnarray}
The surface $\mathcal{C}$ has therefore two components $\mathcal{C}_{\rm in}$ and $\mathcal{C}_{\rm out}$ defined respectively by
$q=q_{\rm in}$ and $q=q_{\rm out}$. Let us analyze now the case of a symmetric head-on collision. 
The condition \eqref{eq:gabpartial=4}
for the continuity of the congruence of normal null geodesics across $\mathcal{C}$
now translates into the couple of equations
\begin{eqnarray}
\Psi'(q_{\rm out})^{2}= {4L^{2}\over q_{\rm out}(q_{\rm out}+1)} \hspace*{1cm}
&\Longrightarrow & \hspace*{1cm} \Psi'(q_{\rm out})=-{2L\over \sqrt{q_{\rm out}(q_{\rm out}+1)}},
\nonumber \\
\Psi'(q_{\rm in})^{2}= {4L^{2}\over q_{\rm in}(q_{\rm in}+1)} \hspace*{1cm}
&\Longrightarrow & \hspace*{1cm} \Psi'(q_{\rm in})={2L\over \sqrt{q_{\rm in}(q_{\rm in}+1)}}.
\end{eqnarray}
The choice of signs is fixed by Eq. \eqref{eq:psi>0interm}. Hence, we have to solve the 
differential equation \eqref{eq:condition1_chordal} with the boundary conditions 
\begin{eqnarray}
\Psi(q_{\rm out})&=& 0,  \label{eq:boundary_cond_nontriv_top1} \\
\Psi(q_{\rm in})&=& 0, \label{eq:boundary_cond_nontriv_top2}\\
\Psi'(q_{\rm out})&=&-{2L\over \sqrt{q_{\rm out}(q_{\rm out}+1)}}, \label{eq:boundary_cond_nontriv_top3} \\
\Psi'(q_{\rm in})&=&{2L\over \sqrt{q_{\rm in}(q_{\rm in}+1)}}. 
\label{eq:boundary_cond_nontriv_top4}
\end{eqnarray}

In the case of the 
solutions found in \cite{arefeva_et_al_1}, it is important to notice that Eq. 
\eqref{eq:boundary_cond_nontriv_top4} is not satisfied at the internal zero of $\Psi(q)$.
Hence, to have a chance of finding trapped surfaces of nontrivial topology we have to be more general.
An important change introduced by the $S^{1}\times S^{D-3}$ 
topology of the trapped surface occurs in the form of the general solution to Eq. \eqref{eq:condition1_chordal}.
Now the region of interest excludes both $q=0$ and $q=\infty$, so there is no reason to set $C_{2}=0$ 
in \eqref{eq:generalsolutiondiffeq} as we did when
assuming that the trapped surface had topology $S^{D-2}$. In fact, $C_{1}$ and $C_{2}$ are determined by Eqs. \eqref{eq:boundary_cond_nontriv_top1} and \eqref{eq:boundary_cond_nontriv_top2} 
\begin{eqnarray}
C_{1}\Phi_{1}(q_{\rm out})+C_{2}\Phi_{2}(q_{\rm out})&=& -\Phi(q_{\rm out})_{\rm RN}, \nonumber \\
C_{1}\Phi_{1}(q_{\rm in})+C_{2}\Phi_{2}(q_{\rm in})&=& -\Phi(q_{\rm in})_{\rm RN}.
\label{eq:C1C2eqs}
\end{eqnarray}
Once these constants are solved in terms of $q_{\rm in}$ and $q_{\rm out}$, we impose the conditions 
\eqref{eq:boundary_cond_nontriv_top3} and \eqref{eq:boundary_cond_nontriv_top4}. 
This provides two algebraic equations that, in principle, are enough to
determine the values of the internal and 
external radii of the trapped surface. 

It is interesting to notice that the resulting function $\Psi(q)$ is independent of 
the mass parameter 
$\mu$. The right-hand side of both equations in \eqref{eq:C1C2eqs} is the sum of two terms, respectively of order 
$\mu^{0}$ and $\mu$. We write $C_{1,2}=C_{1,2}^{(0)}+\mu C_{1,2}^{(1)}$, whereas Eq. \eqref{eq:phiRN} can be recast as
\begin{eqnarray}
\Phi(q)_{\rm RN}
=A_{D}\mu\Phi_{2}(q)+\widetilde{\Phi}(q) \hspace*{0.5cm} \mbox{with} \hspace*{0.5cm} A_{D}=
L{2^{D-3}\sqrt{\pi}\Gamma\left({D\over 2}\right)\over \Gamma\left({D+1\over 2}\right)}\left({G_{N}\over L^{D-3}}
\right).
\end{eqnarray}
Solving the system \eqref{eq:C1C2eqs} order by order in $\mu$ gives $C_{1}^{(1)}=0$, $C_{2}^{(1)}=-A_{D}$. From Eq. 
\eqref{eq:generalsolutiondiffeq} we find that the $\mu$-dependent part in $\Psi(q)$ cancels. 

\begin{figure}
\centerline{\includegraphics[width=2.5in]{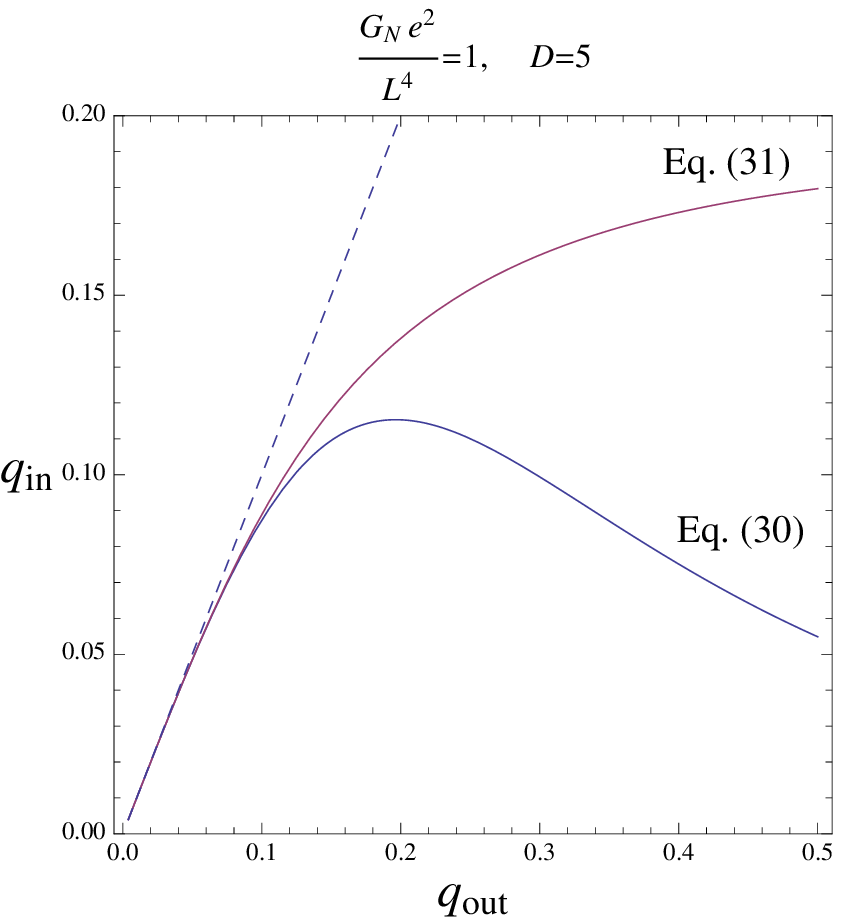}\hspace*{1cm}
\includegraphics[width=2.5in]{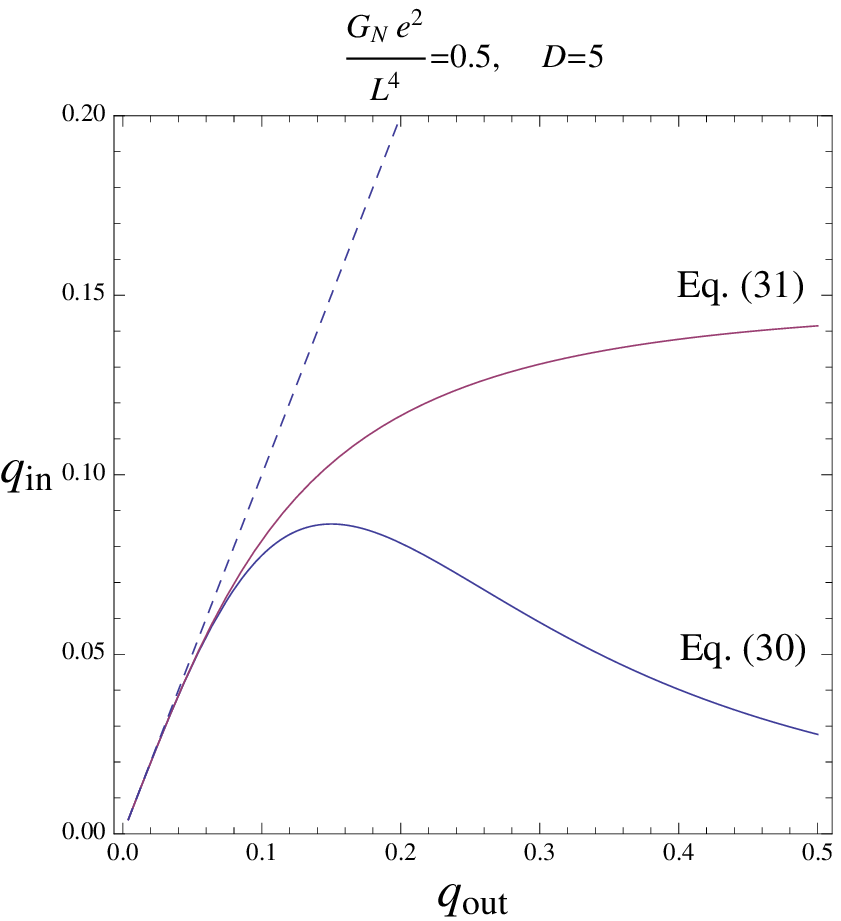}\hspace*{1cm}}
\caption{Plot of the implicit equations \eqref{eq:boundary_cond_nontriv_top3} and \eqref{eq:boundary_cond_nontriv_top4} for 
two values of the charge in $D=5$. The two curves approach each other close to the origin but do not cross. 
The dashed line represents the diagonal $q_{\rm in}=q_{\rm out}$, and shows that both curves lie in the ``physical''
region $q_{\rm out}>q_{\rm in}$.}
\label{fig2_nontrivtop}
\end{figure}

A numerical analysis of the system of equations 
\eqref{eq:boundary_cond_nontriv_top3}-\eqref{eq:boundary_cond_nontriv_top4}, with the values of the constants
found from \eqref{eq:C1C2eqs},
renders no nonvanishing solutions for $q_{\rm in}$ and $q_{\rm out}$. This implies that there are no trapped surface of topology $S^{1}\times S^{D-3}$ formed as the result of the head-on collision of two RN-AdS shock waves. This is pictorially
illustrated in Fig. \ref{fig2_nontrivtop}, where the two curves
defined by Eqs. \eqref{eq:boundary_cond_nontriv_top3} and \eqref{eq:boundary_cond_nontriv_top4} do not cross each other outside the
origin.

\paragraph{Closing remarks}

The study of the problem of collisions of two RN-AdS shock waves presented here shows that the charge parameter $e$ completely 
prevents the formation of marginally trapped surfaces in the region $\{u=0,v\leq 0\}\cup \{u\leq 0,v=0\}$ with topology $S^{D-2}$, independently
of the value of the impact parameter. 
In the case of head-on collisions, we have not found any trapped surface with topology $S^{1}\times S^{D-3}$ in the
same region either. 

These results do not preclude the existence of trapped surfaces
in other regions of the spacetime formed as the result of the collision. In \cite{yoshino_et_al,yoshino_mann} the 
problem of the collision of both Aichelburg-Sexl and 
RN shock waves in flat space was analyzed, looking for trapped surfaces lying in the future light-cone
$\{u= 0,v\geq 0\}\cup\{u\geq 0,v=0\}$. In the case of AdS, the analysis needed to find these trapped surfaces is more
involved than the one required to find the Penrose trapped surface, since the change of
coordinates needed to eliminate the distributional terms in the metric is nontrivial in the ``future ligth-cone'' 
region. As a consequence, the equations
determining the trapped surface are much more complicated and have to be solved numerically, 
even in the case of head-on collisions. This problem is under current investigation.

\section*{Acknowledgments}

We thank Luis \'Alvarez-Gaum\'e, Jos\'e L. F. Barb\'on, and Agust\'{\i}n Sabio Vera for discussions.
Special thanks go to Kerstin E. Kunze for her invaluable assistance with the numerics.
We also thank the Nuclear Physics Group of the University of Salamanca for the use of its
computer cluster. A.D.-V. acknowledges partial support from a
Castilla y Le\'on predoctoral fellowship and Spanish Government Grant FIS2009-07238. The work of M.A.V.-M. has been partially
supported by Spanish Government Grants FPA2009-10612 and FIS2009-07238, Basque Government Grant IT-357-07, and Spanish 
Consolider-Ingenio 2010 Programme CPAN (CSD2007-00042).


\begin{thebibliography}{99}

\bibitem{griffiths}
J. Griffiths, {\it Colliding Plane Waves in General Relativity}, Oxford 1991.

\bibitem{ads/cft}
O.~Aharony, S.~S.~Gubser, J.~M.~Maldacena, H.~Ooguri and Y.~Oz,
  {\it Large N field theories, string theory and gravity,}
  Phys.\ Rept.\  {\bf 323} (2000) 183
  {\tt [hep-th/9905111].}


\bibitem{collisionsAdS}
S.~S.~Gubser, S.~S.~Pufu and A.~Yarom,
  {\it Entropy production in collisions of gravitational shock waves and of heavy ions,}
  Phys.\ Rev.\ D {\bf 78} (2008) 066014
  {\tt [arXiv:0805.1551 [hep-th]].}
\\
S.~S.~Gubser, S.~S.~Pufu and A.~Yarom,
  {\it Off-center collisions in AdS$_{5}$ with applications to multiplicity estimates in heavy-ion collisions,}
  J. High Energy Phys. {\bf 11} (2009) 050
  {\tt [arXiv:0902.4062 [hep-th]].}
  
\bibitem{collisionsAdS_general}
J.~L.~Albacete, Y.~V.~Kovchegov and A.~Taliotis,
  {\it Modeling Heavy Ion Collisions in AdS/CFT,}
  J. High Energy Phys. {\bf 07} (2008) 100
  {\tt [arXiv:0805.2927 [hep-th]].}
\\
J.~L.~Albacete, Y.~V.~Kovchegov and A.~Taliotis,
  {\it Asymmetric Collision of Two Shock Waves in AdS$_5$,}
  J. High Energy Phys. {\bf 05} (2009) 060
  {\tt [arXiv:0902.3046 [hep-th]].}
\\
E.~Avsar, E.~Iancu, L.~McLerran and D.~N.~Triantafyllopoulos,
  {\it Shockwaves and deep inelastic scattering within the gauge/gravity duality,}
  J. High Energy Phys. {\bf 11} (2009) 105
  {\tt [arXiv:0907.4604 [hep-th]].}
\\
A.~Taliotis,
  {\it Heavy Ion Collisions with Transverse Dynamics from Evolving AdS Geometries,}
  J. High Energy Phys. {\bf 09} (2010) 102
  {\tt [arXiv:1004.3500 [hep-th]].}
\\  
E.~Kiritsis and A.~Taliotis,
  {\it Multiplicities from black-hole formation in heavy-ion collisions,}
  {\tt arXiv:1111.1931 [hep-ph].}


\bibitem{AdSGWcolnum}
  D.~Grumiller and P.~Romatschke,
  {\it On the collision of two shock waves in AdS(5),}
  J. High Energy Phys. {\bf 08} (2008) 027
  {\tt [arXiv:0803.3226 [hep-th]].}
\\
P.~M.~Chesler and L.~G.~Yaffe,
  {\it Holography and colliding gravitational shock waves in asymptotically AdS$_5$ spacetime,}
  Phys.\ Rev.\ Lett.\  {\bf 106} (2011) 021601
  {\tt [arXiv:1011.3562 [hep-th]].}
\\
B.~Wu and P.~Romatschke,
  {\it Shock wave collisions in AdS$_{5}$: approximate numerical solutions,}
  Int.\ J.\ Mod.\ Phys.\ {\bf C22} (2011) 1317
  {\tt [arXiv:1108.3715 [hep-th]].}

\bibitem{flat}
R.~Penrose, seminar at Cambridge University, 1974 (unpublished).
\\
P.~D.~D'Eath and P.~N.~Payne,
{\it Gravitational radiation in high speed black hole collisions. 1.
Perturbation treatment of the axisymmetric speed of light collision,}
  Phys.\ Rev.\  {\bf D46} (1992) 658.
  \\
   P.~D.~D'Eath and P.~N.~Payne,
{\it Gravitational radiation in high speed black hole collisions. 2. Reduction
to two independent variables and calculation of the second order news function,}
  Phys.\ Rev.\  {\bf D46} (1992) 675.
  \\
  P.~D.~D'Eath and P.~N.~Payne,
 {\it Gravitational radiation in high speed black hole collisions. 3. Results and
  conclusions,}
  Phys.\ Rev.\  {\bf D46} (1992) 694.
\\
D.~M.~Eardley and S.~B.~Giddings,
{\it Classical black hole production in high-energy collisions,}
  Phys.\ Rev.\  {\bf D66} (2002) 044011
{\tt  [arXiv:gr-qc/0201034].}
\\
E.~Kohlprath and G.~Veneziano,
{\it Black holes from high-energy beam-beam collisions,}
  J. High Energy Phys {\bf 06} (2002) 057
{\tt  [arXiv:gr-qc/0203093].}
\bibitem{LinShuryak}
S.~Lin and E.~Shuryak,
  {\it Grazing Collisions of Gravitational Shock Waves and Entropy Production in Heavy Ion Collision,}
  Phys.\ Rev.\ {\bf D79} (2009) 124015
  {\tt [arXiv:0902.1508 [hep-th]].}

\bibitem{DVVM1}
A.~Due\~nas-Vidal and M.~A.~V\'azquez-Mozo,
  {\it Colliding AdS gravitational shock waves in various dimensions and holography,}
  J. High Energy Phys. {\bf 07} (2010) 021
  {\tt [arXiv:1004.2609 [hep-th]].}
  
\bibitem{AGGSVTVM2}
 L.~\'Alvarez-Gaum\'e, C.~G\'omez, A.~Sabio Vera, A.~Tavanfar and M.~A.~V\'azquez-Mozo,
  {\it Critical formation of trapped surfaces in the collision of gravitational shock waves,}
  J. High Energy Phys. {\bf 02} (2009) 009
  {\tt [arXiv:0811.3969 [hep-th]].}


\bibitem{AS}
P.~C.~Aichelburg and R.~U.~Sexl,
  {\it On the Gravitational field of a massless particle,}
  Gen.\ Rel.\ Grav.\  {\bf 2} (1971) 303.

\bibitem{lousto_sanchez}
C.~O.~Lousto and N.~G.~Sanchez,
  {\it The Curved Shock Wave Space-time Of Ultrarelativistic Charged Particles And Their Scattering,}
  Int.\ J.\ Mod.\ Phys.\ {\bf A5} (1990) 915.

\bibitem{yoshino_mann}
H.~Yoshino and R.~B.~Mann,
  {\it Black hole formation in the head-on collision of ultrarelativistic charges,}
  Phys.\ Rev.\ {\bf D74} (2006) 044003
  {\tt [gr-qc/0605131].}

\bibitem{arefeva_et_al_1}
I.~Y.~Aref'eva, A.~A.~Bagrov and L.~V.~Joukovskaya,
  {\it Critical Trapped Surfaces Formation in the Collision of Ultrarelativistic Charges in (A)dS,}
  J. High Energy Phys. {\bf 03} (2010) 002
  {\tt [arXiv:0909.1294 [hep-th]].}
  
\bibitem{arefeva_et_al_2}
I.~Y.~Aref'eva, A.~A.~Bagrov and E.~O.~Pozdeeva,
  {\it Holographic phase diagram of quark-gluon plasma formed in heavy-ions collisions,}
  {\tt arXiv:1201.6542 [hep-th]}.

\bibitem{hotta_tanaka}
M.~Hotta and M.~Tanaka,
  {\it Shock wave geometry with nonvanishing cosmological constant,}
  Class.\ Quant.\ Grav.\  {\bf 10} (1993) 307.
  
\bibitem{yoshino_nambu}
H.~Yoshino and Y.~Nambu,
  {\it Black hole formation in the grazing collision of high-energy particles,}
  Phys.\ Rev.\ {\bf D67} (2003) 024009
 {\tt [gr-qc/0209003]}.
  
\bibitem{yoshino_et_al}
H.~Yoshino and V.~S.~Rychkov,
  {\it Improved analysis of black hole formation in high-energy particle collisions,}
  Phys.\ Rev.\ {\bf D71} (2005) 104028
   [Erratum-ibid.\ {\bf D77} (2008) 089905]
  {\tt [hep-th/0503171]}.



  

\end{thebibliography}
\end{document}